\documentstyle[multicol,pra,aps,epsf,epsfig]{revtex}
\begin{document}
\title{MASER AND LASER ACTION WITH ONE ATOM}
\author{{\bf N. Nayak}\footnote{nayak@bose.res.in}}
\address{S. N. Bose National Centre for Basic Sciences, 
Block-JD, Sector-3, Salt Lake City, Kolkata-700098, INDIA}
\maketitle
\begin{abstract}
We present a theory which can explain the micromaser as well as its optical counterpart, 
the microlaser, for appropriate values of dissipative parameters. We show that, in 
both the the cases, the cavity radiation fields can have sub-Poissonian photon 
statistics. We further analyse if it is possible to attain a Fock state of the radiation 
field. The microlaser is precluded for such analysics due to the damping of its lasing 
levels making transitions at optical frequencies. Hence, we focus our attention on the 
micromaser and our exact simulation of the dynamics shows that it is not possible to 
generate a Fock state of the cavity radiation field.
\end{abstract}
\begin{multicols}{2}
\section {Introduction}
The subject of one-atom maser and its optical counterpart, the one-atom laser, has
generated extensive interest after the recent experimental demonstrationa that they 
are capable of generating nonclassical states of the radiation fields [1-3].
 In the one-atom maser experiment [1], two-level $^{85}Rb$ atoms in their upper 
Rydberg states of the transitions $63^{2}p_{3/2}\rightarrow 61^{2}d_{3/2} 
(21.5065~GHz)$ and $63^{2}p_{3/2}\rightarrow 61^{2}d_{5/2} (21.456~GHz)$ are pumped 
into a microwave cavity at such a rate that, at most, one atom is present there at 
any time. So, there can be one of the two situations in the cavity: only one atom is 
present in the cavity or the cavity is empty of any atom. The cavity is tuned to 
one of the above two transitions in individual experiments and the sub-Poissonian 
nature of the cavity radiation field having variance less than that for a coherent 
state field has been inferred. It may be noted here that a well stabilized 
conventional laser can generate radiation fields that can at best be close to a 
coherent state field only. Sub-Poissonian radiation fields have photon distribution 
functions narrower than that for a coherent state field which has a Poissonian 
photon statistics.

Whereas in the one-atom laser experiment [2], the pump rate of two-level $138 Ba$ 
atoms in their upper states into a resonant optical cavity gives a stream of an 
average number of atoms present there satisfying the condition $\langle N\rangle\leq 1.0$. 
However, in order to have a sub-Poissonian radiation field, 
an uniform atom-field interaction independent of cavity mode strucure is essential. The 
technique adopted by An el al [3] to improve their earlier setup [2] provides such a 
situation. Single-atom events in this cavity would be suitable for generating sub-Poissonian 
field in the optical regime since such arrangements in the microwave regime 
produced nonclassical fields [1].


In the following, we describe a unified formalism which shows that the one-atom 
maser (micromaser) as well as the one-atom laser (microlaser) cavity fields can have 
sub-Poissonian photon statistics. Our unified theory also explains the conventional 
laser dynamics. Further, the hope is that such sub-Possonian radiation fields would 
further shrink to bosonic Fock states. Our analysis below shows that this unified 
theory is not suitable for looking into these possibilities. Hence, we examine the 
possibilities of generating Fock states of the micromaser cavity field by an exact 
numerical simulation of the dynamics. In the case of microlaser, the damping of the 
atomic levels precludes such possibilities.\\ \\ 
\noindent
\section{The formalism}
We assume that atoms arrive individually at the cavity with an average interval 
$\bar{t}_{c}=1/R$ where $R$ is the flux rate of atoms. We have 
$t_{c}=\tau + t_{cav}$ where $\tau$ is the interaction time, fixed for every atom, 
and $t_{cav}$ is the random time lapse between one atom leaving and successive atom 
entering the cavity. $\bar{t}_{c}$ is the average of $t_c$ taken over a Poissonian 
distribution in time of incoming atoms. The cavity field evolves by this repititive
 dynamics from near vacuum as thermal photons in the optical cavity are almost 
nonexitent. Thus, during $\tau$ , we have to solve the equation of motion by taking 
into account the cavity dissipation as well as atomic damping, important at optical 
frequencies. Thus, we have
 \begin{eqnarray}
\dot{\rho} &=& -i[H,\rho ]-\kappa (a^{\dagger}a\rho -2a\rho a^{\dagger}+
\rho a^{\dagger}a)
\nonumber \\
& & -\gamma (S^+ S^- \rho -2S^- \rho S^+ +\rho S^+ S^-)
\end{eqnarray}
where $H$ is the Jaynes-Cummings Hamiltonian [6] and $\kappa$ and $\gamma$
are the cavity-mode and atomic decay constants respectively. $a$
is the photon annihilation operator and $S^+$ and $S^-$ are the Pauli
pseudo-spin operators for the two-level system.
During $t_{cav}$, the cavity field evolves under its own dynamics and is represented by 
\begin{eqnarray}
\dot{\rho} &=& -\kappa(1+\bar n_{th}) (a^{\dagger}a\rho -2a\rho a^{\dagger}+
\rho a^{\dagger}a)
\nonumber \\
 & & -\kappa \bar n_{th} (aa^{\dagger}\rho -2a^{\dagger}\rho a +
\rho aa^{\dagger})
\end{eqnarray}
The method for obtaining a coarse-grained time 
derivative for the photon number distribution $P_n =<n\vert \rho \vert n>$ is given
in detail in Ref. [4]. The steady-state photon statistics is then
\begin{equation}
P_{n}=P_{0}\prod_{m=1}^{n}v_{m}
\end{equation}
and $P_{0}$ is obtained from the normalisation $\sum^{\infty}_{n=0}P_{n}=1$.
The $v_{n}$ is given by a continued fractions invoving the system parameters
 $N=R/2\kappa$, the number of atoms passing through the cavity in a 
photon lifetime, $\kappa /g$ and $\gamma /g$ where $g$ is the atom-field coupling 
constant. Once $P_n$ is obtained, we can get the laser intensity, proportional to 
$\langle n\rangle$, and its variance. {\bf This represents micromaser or microlaser
photon statistics depending upon the relative values of the dissipative parameters 
and the coupling constatnt.} We introduce another paprameter, the pump parameter
$D=\sqrt{N}g\tau $ useful for the description of laser characteristics. 
It is interesting to note that we can recover the conventional laser photon 
statistics from Eq. (2) by an appropriate choice of the above parameters. \\ \\
\noindent
\section{ One-atom maser dynamics}
This involves the Rydberg levels of the active atoms having spontaneous lifetime 
$t_s =1/2\gamma$ in the order of a fraction of a second. Whereas their flight 
time $\tau$ through the cavity is of the order of microseconds.
Hence we can safely put $\gamma =0$ in the Eq. (1). We find from our formalism 
that the normalized variance defined by
\begin{equation}
 v=\sqrt{(\langle n^2\rangle -\langle n\rangle ^2)/\langle n\rangle}
\end{equation}
is $0.5522$ which is very close to the experimental finding $v=0.5477$ 
in Ref. 1a near $N=30$ and values of the other parameters reported there.
Thus we find that our formalism describes the micromaser dynamics quite 
accurately. It may be noted here that $v~=~1$ for coherent state radiation fields. 
Thus, this novel device does generate radiation fields with sub-Poissoinian 
photon statistics.  \\ \\
\noindent
\section{One-atom laser dynamics}
The lasing frequency here is in optical regime and, hence, the atomic lifetimes 
in these cases are short. Hence, we have to take $\gamma\ne 0$ for the microlaser 
dynamics. For the typical parameters in the experiments in Ref. 3, we find that 
$\gamma /g =0.1$. From our photon statitics in Eq. (2), we obtain the laser intensity 
proportional to $\langle n\rangle$ as a function of the pump parameter $D$. 
The structures in $\langle n\rangle$ as $D$ is varied for fixed $N$ 
reflect the characteristics of the Jaynes-Cummings interaction[6]. Soon after
the threshold is attained at about $D=1.0$, the photon number rises sharply.
The reason is as follows. The field is almost in vacuum before the very first
atom enters the cavity. Thus the atoms initially in their upper states 
contribute varying fractions of their energies to the cavity and at $n=N$ 
and $D=\pi /2$ the atoms
get completely inverted. Thus $\langle n\rangle$ is peaked at about $D=1.6$ depending on
$\kappa$, $\gamma$ and $N$. For higher $\kappa$ and $\gamma$, the peak moves
slightly towards higher $D$ due to threshold being attained at higher $D$.
We further find that $<n>=0$ around $D=$31.4, 62.8, 94.2,.... giving 
$g\tau =\pi$, $2\pi$, $3\pi$,....  respectively. At such values of $g\tau$, 
the atom absorbs the photon it has emitted before leaving the cavity.\\
 \indent
 The variance of the cavity field $v$ increases sharply at about $D=1.6$ where 
$\langle n\rangle$ is also peaked. We find that near this value of $D$, the $P_n$ is 
doubly peaked at $n=0$ and $n\simeq N$ and this increases the variance in photon number. 
However, for slightly higher value of $D$, the cavity field is highly sub-Poissonian in 
nature. It also appears for further higher values of $D$. But, $\langle n\rangle$ is very 
small for such values of $D$ due to increase in the interaction time $\tau $ which
 increases the infuence of atomic as well as cavity reservoirs on the
 atom-field interaction. It may be noted here that these characteristics are 
 clearly different from that of conventional lasers.\\ \\
\noindent
\section{Can there be a Fock state of the radiation field?}
We see that the one-atom dynamics is capable of narrowing the photon 
distribution function of the cavity radiation field compared to that for a 
coherent state field. This raises the question whether the distribution can be 
further reduced to a number state of the radiation field. The reason why such 
possibilities may arise is due to the fact that the reservoir influences are 
minimal in these dynamics. In fact, the analysis in Ref. [4] indicates this 
possibility in the micromaser dynamics when the condition 
\begin{eqnarray}
f(n) &=& -2n\bar n_{th}-2N\sin^2(\sqrt{n}g\tau)   \nonumber \\
     & & \times\exp[-\gamma-(2n-1)\kappa]\tau~=~0
\end{eqnarray}
is satisfied. Clearly $f(n)~=~0$ only when $\bar n_{th} =0$ with $g\tau=\pi/2$. 
The later condition can be met by an adjustment of the atomic velocity with the 
help of the velocity selector in the micromaser setup [1]. But the other condition 
that the cavity temperature $T~=~0~K$ obviously cannot be satisfied.
In the micromaser experiment [1b], the quality factore of 
the cavity is $Q=3.4\times 10^{9}$ and its temperature is maintained at $T=0.3~K$ 
which gives thermal photons $\bar n_{th}=0.033$. The Rydberg level involved 
in the atomic transitions have lifetimes of the order of a second which is 
much longer compared to its flight time of about $\tau=40~\mu s$ through the cavity. 
These values of the parameters governing the dissipation dynamics make their 
influences very small. In the case of one-atom laser, the lifetimes 
of atomic levels at optical frequencies are short which makes the influence 
of atomic reservoir non-negligible. Hence, even though the thermal photons 
$\bar n_{th}=0$ in the optical cavity, the one-atom laser [2,3] would not be a
right choice for this study. Thus, the micromaser turns out to be a 
suitable system to look into the possibilities of generating a number state 
of the cavity radiation field. \\ \\
\indent
We then have to take into account the exact infuences of the reservoirs to acertain 
if a Fock state of the cavity field can actually be generated. 
Hence, instead of Eq. (1), we have to deal with \\ 
 \begin{eqnarray}
\dot{\rho} &=& -i[H,\rho ]-\kappa(1+\bar n_{th}) (a^{\dagger}a\rho -2a\rho a^{\dagger}+
\rho a^{\dagger}a)
\nonumber \\
 & & -\kappa \bar n_{th} (aa^{\dagger}\rho -2a^{\dagger}\rho a +
\rho aa^{\dagger})
\end{eqnarray}
whenever a atom is present in the cavity. As mentioned earlier, a atom takes 
a time $\tau$ to pass through the cavity. These atomic events are seperated 
by random durations, $t_{cav}$, during which the cavity evolves under its own 
dynamics.  Hence we set $H=0$ during $t_{cav}$. Processes like these stomic 
events seperated by random intervals are known as Poisson processes 
in literature encountered in various branches of physics, for example, 
radioactive materials emitting alpha particles. A sequence of durations of such 
processes can be obtained from uniform deviates, also called random numbers, $x$ 
generated using a computer such that $0~<~x~<1$, and then by using the 
relation [7]
\begin{equation}
t_R=-\mu\ln(x)
\end{equation}
where $t_R=t_{cav}+\tau$. $\mu~=~1/R$ where $R$ is the flux rate of atoms.  \\
\indent
We have carried out numerical simulation of the dynamics with the data taken 
from the experimental arrangements [1] in which $g=39~kHz$ and the $\tau =40~\mu s$ 
was one of the atom-field interaction times. In this case $g\tau = 1.56$ 
which produces  Fock states of n photons where n satisfies 
$\sin g\tau\sqrt{n+1}=0$ in an ideal cavity $(Q=\infty)$. Since the experimantal 
arrangements are close to ideal situation, it was hoped that such Fock states could be 
attained experimentally. Indeed, such results have been reported in Ref. 1b. However, 
our numerical simulations does not confirm these conclusions. Instead, it gives photon 
fields with very narrow distribution functions (sub-Poissoninan) centred about n. Figs. 
1 and 2 display distribution function $P(n)$ narrowly centred about $n=14$.\\ 


\begin{figure}
\centerline{\epsfig{file=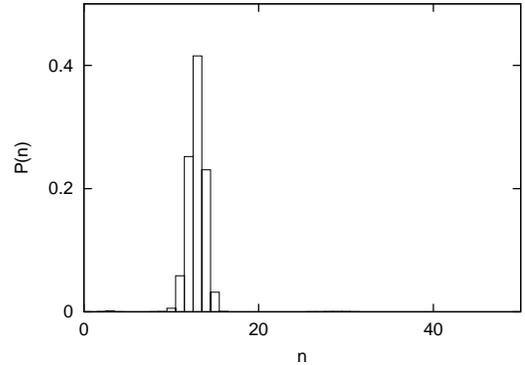,height=2.0in}}
{\caption{
Cavity photon distribution function at the exit of the 7000th atom.
}}
\end{figure}

\begin{figure}
\begin{center}
\centerline{\epsfig{file=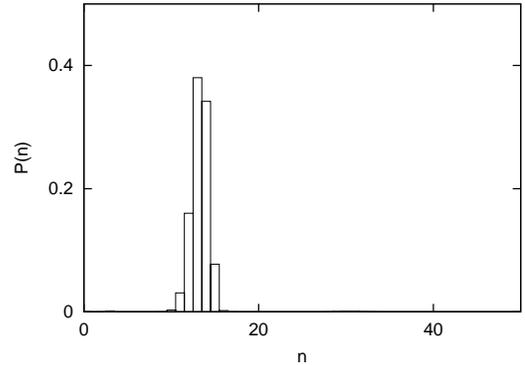,height=2.0in}}
{\caption{
P(n) vs n at the moment of the 9000th atom leaving the cavity.
}}
\end{center}
\end{figure}


The reason for these results is simple. The cavity dissipation, although very small, 
effects the coherent atoim-field interaction and moreover the randomness in $t_{cav}$ 
makes the photon distribution function fluctuate all the time centred about n in 
addition to making it broader.\\ 
\indent
In this experiment [1], the atoms coming out of the cavity are subjected to 
measuments from which state of the cavity field is inferred. The atoms enter 
the cavity in the upper $\vert a\rangle$ of the two states $\vert a\rangle$ and 
$\vert b\rangle$. The exiting atom is, in general, in a state 
\begin{equation}
\vert\psi\rangle = a\vert a\rangle + b\vert b \rangle
\end{equation}
with $p_a=\vert a\vert ^2$ and $p_b=\vert b\vert ^2$ are the probabilities of the 
atom being in the states $\vert a\rangle$ and $\vert b\rangle$ respectively.
According to the Copenhagen interpretation of quantum mechanics [8], this 
wave function {\it collapses} (or is projected) to either $\vert a\rangle$ or 
$\vert b\rangle$ the moment a measurement is made on it. Due to this inherent 
nature of quantum mechanics, a noise is associated with the measurement 
which is know as {\it quantum projection noise} [9]. We define the projection 
operator $J=\vert a\rangle\langle a\vert$. The variance in its measurement 
is given by
\begin{equation}
(\Delta J)^2=\langle J^2\rangle - \langle J\rangle^2=p_a(1-p_a)
\end{equation}
We find that $(\Delta J)^2=0$ only when $p_a=1~or~0$. For the generation of a 
Fock state, it is necessary that the atom should leave the cavity unchanged in its upper 
state [4]. Hence, for such a situation we must have $p_a=1$ in which case 
$(\Delta J)^2$ should be 0. We find from our numerical simulations 
that that $p_a\equiv P(a)$ is mostly about 0.8 [Fig. 3] and, hence, 
$(\Delta J)^2\ne 0$ always. This obviously indicates that the cavity 
field is in a linear superposition of Fock states giving a photon distribution 
function with the variance, defined in Eq. (3), $v~>~0$ (For a Fock state $v=0$). 
Indeed, we find that the $v$ is about $0.5$ in our calculations, presented in Fig. 4, 
indicating a sub-Poissonian nature of the cavity field. By itself, it carries a 
signature of quantum mechanics. \\ 

\end{multicols}
\begin{figure}
\begin{center}
\centerline{\epsfig{file=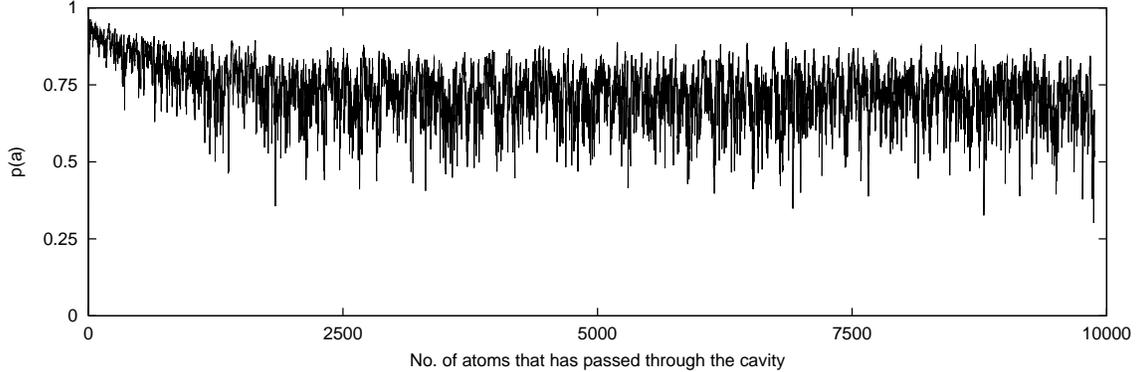,height=2.0in}}
{\caption{Population of the upper state of the individual atoms at the exit from 
the cavity.}}
\end{center}
\end{figure}

\begin{multicols}{2}

We further notice in Fig. 4 that there are small fluctuations in $v$ due to the 
fluctuations in $P(n)$ [Figs 1 and 2]. Also, $v$ is nowhere near $0$ in Fig. 4.\\ 
\begin{figure}
\begin{center}
\centerline{\epsfig{file=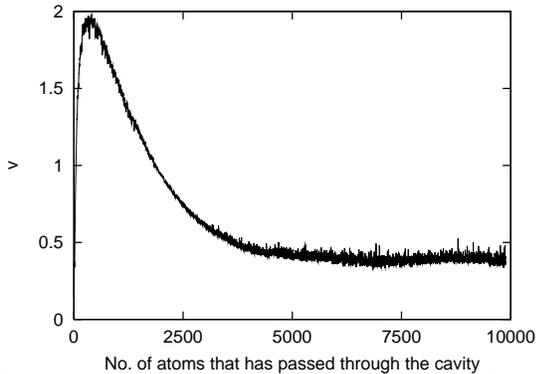,height=2.0in}}
{\caption{
Fluctuations in $v$ at the exit of successive atoms from the cavity.
}}
\end{center}
\end{figure}
The steady-state photon statistics in Eq. (2) gives $v=0.597$ which lies within the 
range of fluctuations in Fig. 4. This is due to the fact that the coarse-graining 
process involved in obtaining steady-state results neglects these small fluctuations 
(A detailed discussion can be found in Ref. [5]). Our exact numerical simulations, 
however, shows that they are crucial in the generation of a Fock state. We have carried 
out simulation until about 10000 atoms passed through the cavity and carrying out the 
simulations any further would only be a repitition of the above fluctuations. \\ \\
{\bf References:}\\
\noindent
[1a] G. Rempe et al, Phys. Rev. Lett. {\bf 64}, 2783 (1990).\\
\noindent
[1b] B. T. H. Varcoe et al, Nature, {\bf 403}, 743 (2000). \\
\noindent
[2] K. An et al, Phys. Rev. Lett. {\bf 73}, 3375 (1994).\\
\noindent
[3] K. An, R. R. Dasari, and M. S. Feld, Opt.lett. {\bf 22}, 1500 (1997).\\
\noindent
[4] N. Nayak, Opt.Commun. {\bf 118}, 114 (1995).\\
\noindent
[5] N. Nayak, J. Opt. Soc. Am. B {\bf 13}, 2099 (1996).\\
\noindent
[6] E. T. Jaynes and F. W. Cummings, Proc.IEEE {\bf 51}, 89 (1963).\\
\noindent
[7] D. E. Kunth, {\it The Art of Computer Programming} (addison-Wesley, Reading, 
Mass. 1981), Vol.2.\\
\noindent
[8] Max Jammer, {\it The Conceptual Developement of Quantum Mechanics} (McGraw-Hill, 
1966), Ch.7. \\
\noindent
[9] W. M. Itano {\it et al}, Phys. Rev.  A {\bf 47}, 3554 (1993).
\end{multicols}
\end{document}